\documentclass[aps,
               reprint,
               prb,
               twocolumn,
               superscriptaddress]{revtex4-2}
\usepackage{bm}
\usepackage{graphicx}
\usepackage{graphics}
\usepackage{amsmath}
\usepackage{amsfonts}
\usepackage{amssymb}
\usepackage{makeidx}
\usepackage[colorlinks=true,citecolor=blue,linkcolor=blue,linktocpage=true,pagebackref=false]{hyperref}
\hypersetup{colorlinks=true,citecolor=blue,linkcolor=blue,filecolor=blue,urlcolor=blue}
\usepackage{color,soul} 
\usepackage{subfigure}
\usepackage{placeins}
\usepackage{xcolor}
\usepackage{caption}
\usepackage{siunitx}
\usepackage{tabularx}
\usepackage{array}
\usepackage{ragged2e}

\graphicspath{{figures/}} 
\begin{document}

\newcommand{\be}   {\begin{equation}}
\newcommand{\ee}   {\end{equation}}
\newcommand{\ba}   {\begin{eqnarray}}
\newcommand{\ea}   {\end{eqnarray}}
\newcommand{\ve}   {\varepsilon}
\newcommand{\Dis}  {\mbox{\scriptsize dis}}

\newcommand{\state} {\mbox{\scriptsize state}}
\newcommand{\band} {\mbox{\scriptsize band}}
\newcolumntype{Y}{>{\centering\arraybackslash}X}

\title{Slide and Twist: Manipulating Polarization in Multilayer Hexagonal Boron-Nitride}


\author{Sanber Vizcaya, Felipe Perez Riffo, Juan M. Florez, Eric Su\'arez Morell}

\affiliation{Grupo de Simulaciones, Departamento de F\'isica, Universidad T\'ecnica Federico Santa Mar\'ia, Valpara\'iso 2390123, Chile}

\date{\today}
\begin{abstract}

This study explores the world of across-layer sliding ferroelectricity in multilayer hexagonal boron nitride (hBN) and gallium nitride (hGaN), aiming to control out-of-plane polarization. By investigating the effects of sliding single or dual layers in various hBN stacking configurations, we uncover methods for reversing polarization with energy barriers between 5 and 30 meV/f.u., making these methods experimentally viable. Our results show that single-interface sliding is more energetically favorable, with lower barriers compared to multiple interfaces. Certain pathways reveal stable polarization plateaus, where polarization remains constant during specific sliding phases, promising robust polarization control.
Moreover, rotated multilayer structures maintain consistent net out-of-plane polarization across different rotation angles. In trilayer ABT structures, rotating the top layer and sliding the bottom layer can reverse polarization, expanding device design possibilities. While the primary focus is on hBN, similar phenomena in hGaN suggest broader applicability for this class of polar materials. The identified energy barriers support the feasibility of fabricating devices based on these multilayer structures.

\vspace{10pt}


\end{abstract}

\maketitle

\section{Introduction}
\label{sec:introduction}
Research on two-dimensional ferroelectric (FE) materials presents an exciting avenue for discovering new layered materials with potential applications in technologies like non-volatile memories, field-effect transistors (FETs), and cryogenic memory devices \cite{Xu2022,Sen2024}. A key challenge in this area is achieving precise control over polarization, which can be done, e.g., by sliding between two bi-stable polar states in one of the material's layers \cite{Sui2024}. This process breaks the inversion and/or reflection symmetries along a horizontal plane, which is the case in 2D hexagonal-like materials with vertical polarization \cite{zhang_vertical_2023}, resulting principally in an imbalance in charge transfer, leading to changes in such electric polarization.

This so-called across-layer sliding ferroelectricity (ALSF) stems from the asymmetry of next-neighbor inter-layer couplings. It offers methods of controlling polarization like using low energetic electron-beam illumination or scanning the surface with a biased tip. Those methods are an alternative to external field-driven mechanisms, which might require large energies\cite{yang_across-layer_2023}. Changes in electric polarization associated with charge transfer are usually small and generally appear at very low temperatures, limiting their applicability \cite{Xu2022}. However, the switching behavior may still be robust at higher temperatures, suggesting that key factors for a typical ferroelectric model are still missing in these systems. Therefore, new pathways such as multilayer-sliding are needed to improve the FE results of typical charge transfer-based ALSF \cite{yang_across-layer_2023, yasuda_ultrafast_2024}.

Different experiments have shown that slidetronics is within reach. For instance, Meng et al. reported the reversion of the polarization in the transition from a cyclic stacking (...ABCABC...) to an anti-cyclic one (...CBACBA...) in 3R MoS$_2$ trilayers \cite{RePEc}. Sui et al. visualized the enhanced polarization switching in $\gamma$-InSe, which originated from microstructure modifications such as stacking fault elimination and subtle rhombohedral distortions due to continuous interlayer sliding \cite{sui_sliding_2023}. In addition, Stern et al. showed that reversible polarization switching coupled to lateral sliding in the h-BN bilayer is achieved by scanning a biased tip above the surface \cite{vizner_stern_interfacial_2021}. Hexagonal nitrides are particularly interesting due to their large out-of-plane polarization (OOP), small energy barriers between stackings, and easy integration into heterostructures with graphene or transition metal dichalcogenides\cite{ViznerStern2024,vizner_stern_interfacial_2021,zhang_vertical_2023,CORTES2023111086}. In the case of these last two nitrides, bilayer structures have displayed an OOP of $\sim$ 2.0 pC / m in hBN with stacking of AB (BA) \cite{Yasuda,YANG}, and polarizations of $6 \text{ pC/m}$ were obtained for hGaN \cite{Wang2023}. Also, hBN trilayers were recently shown to have checkerboard-shaped triangular domains. The polarization of such domains could be customized by introducing selected twisting between layers \cite{CORTES2023111086}, mimicking the rotated hBN bilayer, where triangular domains with excess/defect charge were shown to follow moire patterns and unit cell reconstructions \cite{Hennighausen_2021}. In addition, tunneling junctions based on twisted hBN bilayer showed robust symmetric FE domains that could be controlled by mechanical stress \cite{lv_spatially_2022}.

One big question in out-of-plane stacking ferroelectricity (SFE) is whether symmetry is the predominant factor defining the resulting FE. Recently, a geometric approach clarified the Berry phase origin behind SFE and established a general condition for robust out-of-plane polarization, which goes beyond current symmetry-based understanding. In the SFE bilayer, it was shown that the Berry phase behind non-zero $P_{z}$ is due to a correlation between the interlayer potential, an effective staggered sublattice potential, and the hopping between adjacent AB sites \cite{Zhou}. The dominant contribution to the Berry phase seems to come from the vicinity of the symmetry points K and Q in the AB-stacked honeycomb bilayer and bilayer $T_d$ TMD, respectively\cite{Zhou,yu_distinct_2023}. This has not been tested beyond bilayers, which motivates the use of sliding to break the symmetry, e.g., beyond the horizontal mirror breaking due to AB stacking creating topologies that disturb the asymmetric interlayer coupling. This would be of great interest to understand the across-layer sliding ferroelectricity.

ALSF studies in multilayers are relatively new as structural phase transitions of crystals are challenging to control owing to the energy cost of breaking dense solid bonds. However, electric field switching of stacking in vdW multilayers has opened the door to new possibilities \cite{ViznerStern2024,yasuda_ultrafast_2024,Sui2024}. The multiple paths for sliding in honeycomb layers define the so-called polytypes \cite{ViznerStern2024}, which increase with the number of layers. Bilayers are relatively simpler to analyze \cite{zhang_vertical_2023}, three/four layers are a bit more challenging \cite{CORTES2023111086,yang_across-layer_2023}, and to the best of our knowledge, there have been no studies of ALSF in tetralayered hBN and some questions remain in the trilayered case. The characterization of four/tri-layer could help identify the challenges of harnessing the switching mechanisms for rapid, local, and practical ferroic devices because two layers could slide between/on two/one exterior ones that, e.g., could be pinned to a substrate or studied by using conductive atomic force microscopy, respectively \cite{lv_spatially_2022}. This number of layers would also be interesting considering the elimination of stack faults and the distortion due to intralayer compression and continuous interlayer sliding \cite{sui_sliding_2023}.

In general, the electronic properties of systems with 1-4 layers are different from bulk 3D crystals\cite{Wickramaratne2018}.  Young's modulus converges to bulk values when the number of layers exceeds eight \cite{Zhou2024}. The OOP dielectric constant reaches bulk values above five layers, while the in-plane component does not change with the number of layers \cite{ChunminMa}. In light of these results, the trilayer/tetralayer nitrides are still within the 2D world and emerge as very interesting for an ALSF study. In this work, we studied layered hBN and hGaN, considering varied sliding paths between representative stacking and specific rotations. By using the modern theory of polarization and semi-classical approaches, we calculate the electric polarization $P_{z}$ for different layer sliding paths and twisted layered nitrides, using, in this last case, Tight Binding (TB) modeling. We show that rotation of the third and fourth layers for specific stacking provides us with finite polarizations that could be used for sliding-tuned FE switching. Our study suggests the possibility of achieving a net polarization in multilayer structures rotating such layers, which differs from what occurs in freestanding rotated bilayers, where the net polarization is zero. We obtain polar states of $\simeq \ 4.5\ { pC/m}$ in four-layer ABAB structures with two rotated layers and $\simeq \ 3.0\ { pC/m}$ in three-layer structures when the outer layer is rotated.

This paper is organized as follows: In Section \ref{Methodology}, we explain the methodology, the parameters used for DFT and TB calculations, and how the relaxations of large systems were carried out. We show and discuss the results in Section \ref{Results}; finally, in Section \ref{sec:conclusion}, we draw some conclusions.

\begin{figure}[t]
\centering
\includegraphics[width=\columnwidth]{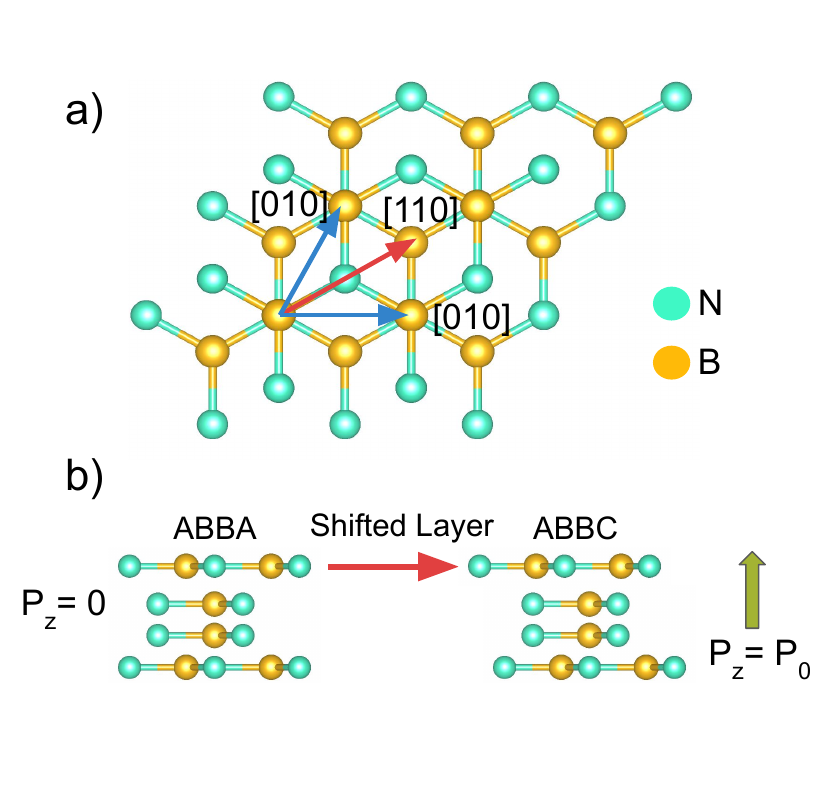}  
\captionsetup{justification=raggedright, singlelinecheck=false, font=small}
\vspace{-1.0cm}
\caption{\justifying (a) Top view of a bilayer structure for a [1,1,0] sliding direction. (b) Lateral view of a tetralayer structure with the top/fourth layer shifted from ABBA to ABBC respect the third one; there exists an effective OOP polarization.}
\label{fig1_sliding}
\end{figure}

\section{Methodology}
\label{Methodology}
\label{DFT}

We performed density functional theory (DFT) calculations by using the Vienna Ab-initio Simulation Package (VASP 6.2.0) \cite{PLIMPTON19951,Kresse}. We employed the Generalized Gradient Approximation with the Perdew–Burke–Ernzerhof functional for exchange-correlation. A cutoff energy of 520 eV was used, with a unit cell including a 20 Å vacuum along the z-axis to prevent image-related issues. A k-point mesh of 12 x 12 x 1 was used for structural relaxations, while a 41 x 41 x 1 was used for static calculations. For twisted multilayers, a 7 x 7 x 1 grid was used. Structural relaxations were performed by ensuring that the forces on all atoms were below 0.005 eV/Å. The PBE-D2 functional proposed by Grimme was also used to account for dispersion forces, and the electric polarization was calculated using the Berry phase within the modern theory of polarization scope. The classical polarizations obtained from a Bader-charges approach \cite{Grimme2006, King-Smith1993} were also calculated for comparison/discussion purposes. The energy barriers that characterize the sliding between different stacking were calculated using the Nudge Elastic Band (NEB) method \cite{Henkelman,Bader}.

\label{Tight Binding Model}
Twisted multilayers present large unit cells, which are not suitable for DFT calculations. We use semiclassical methods to calculate the FE in those cases instead, with a local charges approximation adapted to a TB modeling of single orbitals. This method reproduces several trends observed in the Bader-charges approach and several characteristics obtained using DFT in smaller systems. We consider first-neighbor hopping within the layers, with an on-site energy of +/-4 eV, and an interlayer hopping specific to each atom: between Boron-Boron of $t_{BB}$ = 0.7 eV, Nitrogen-Nitrogen of $t_{NN}$ = 0.15 eV, and between Boron-Nitrogen of $t_{BN}$ = 0.3 eV. This hopping follows an exponential decay according to the distances given by the following expression: $t_{XY}(r)=t_{XY} \exp[-\alpha(r-d)]$, where $\alpha=4.4$ and $d=3.35$ \AA{} \cite{Guinea2021}. Using TB, we calculate the polarization semi-classically and use a rescaling factor ($\sim$ 1.5) to match both results. The TB only uses one p$_z$ orbital per atom, underestimating the result.


\label{sec:electronic-calculations}
To determine the relaxed atomic positions for twisted h-BN multilayer at a given twist angle $\theta$, we employed the Large-scale Atomic/Molecular Massively Parallel Simulator (LAMMPS) \cite{LAMMPS2022}. This method ensures accurate modeling by incorporating bonded and non-bonded interactions, which is crucial to account for different distributions of stacking within the supercells, cell reconstruction, and energetically favorable structural stabilization. For non-bonded interlayer interactions, we used a registry-dependent interlayer potential following the parametrization of Ouyang et al. \cite{Ouyang2018}. This potential is tailored for graphene/hBN heterostructures \cite{Leven2014,Leven2016,Maaravi2017} and has a cut-off distance of 16 \,\AA\, to account for all significant forces. For bonded intralayer interactions, we applied the Tersoff potential \cite{Tersoff1988,Kinaci2012}, which describes covalent bonds in materials such as boron nitride. Finally, the energy minimization of the layered structure was performed with an energy tolerance of $10^{-10} \, \text{eV/atom}$ \cite{Mostofi_2020}.

\section{Results and Discussion}
\label{Results}
\subsection{Tetralayer}
\label{TetraLayers}

Figure \ref{fig1_sliding} illustrates our first sliding analysis, which is based on using the top layer to sense-like the FE response of the lower trilayer as the tetralayer is formed with its top layer shifted along different stacking. This intuitive approach would mimic, for example, what could be done using atomic force microscopy \cite{lv_spatially_2022,vizner_stern_interfacial_2021}. The polarization difference is taken with respect to the centrosymmetric structure ABBA that has $m_{z}$ symmetry, which prohibits an effective OOP polarization. In contrast, all other stacking belongs to the symmetry group P$_3$ with only $C_{3z}$ rotations.

Following the whole path in Figure \ref{fig1_sliding}, the polarization changes continuously, ranging from 1.85 to -5.30 pC/m. Two energy zones are characteristic, one for a great barrier of up $\sim30\text{meV/f.u.}$, covering half of the migrations, and one with smaller barriers ranging between 5 and 10 meV/f.u., that is, ABAC-ABBC-ABBA-ABCA and ABAC-ABAB-ABCB-ABCA, respectively. The largest barrier might be attributed to specific sections where direct stacking occurs (atoms of the same type positioned directly on each other), making them less stable energetically. Low-energy barriers are promoted when short-range electronic repulsion is the primary contributor and only one sliding interface exists. Cortés et al. showed that for h-BN trilayers that follow such a trend, the energy barrier seems additive for each interface \cite{CORTES2023111086,RePEc}. From the polarization viewpoint, in the vicinity of ABAC, the polarization change is positive; however, it is negative anywhere else. This could be attributed to preparing the first two layers in an AB stacking. In what follows, we discuss how to change that polarization alignment and values.


\begin{figure}[t]
\centering
\includegraphics[width=1.0\columnwidth]{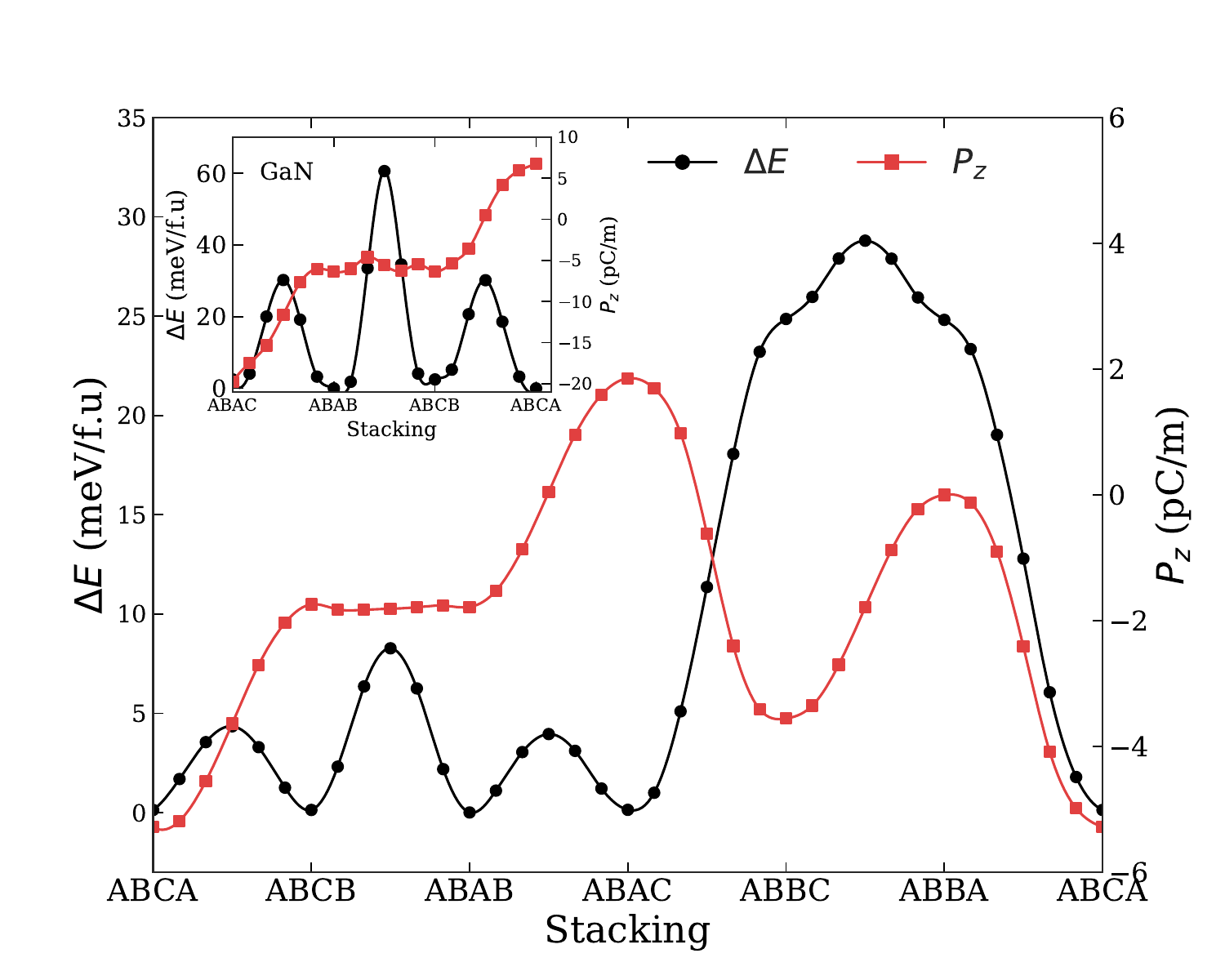}
\captionsetup{justification=raggedright, singlelinecheck=false, font=small}
\vspace{-0.1cm}
\caption{\justifying Relative energy-barriers and corresponding OOP polarization for sliding paths connecting different hBN tetralayer's stacking (letters label from bottom (left) to top-layer (right)). Inset: reverse path of first half of hBN's one, containing the plateau, but using GaN instead of BN.}
\label{fig2_sliding}
\end{figure}

\begin{figure}[t]
\centering
\includegraphics[width=1.0\columnwidth]{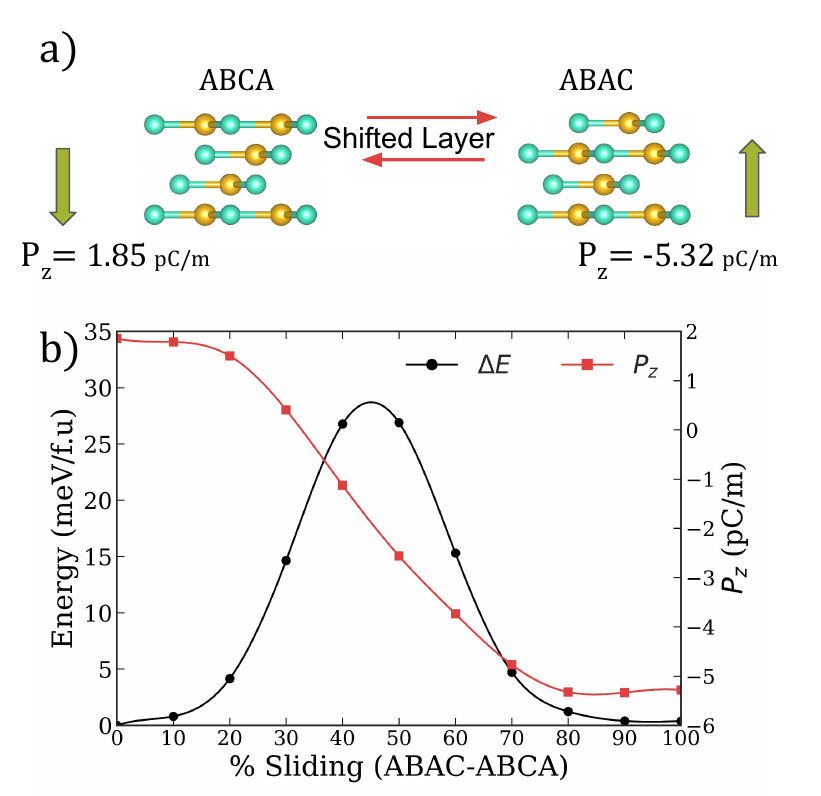}
\captionsetup{justification=raggedright, singlelinecheck=false, font=small}
\vspace{-0.1cm}
\caption{\justifying (a) Lateral view of double sliding in hBN tetralayer, where third and fourth layers are both shifted at each interpolation step between end-stacking; OOP polarization for those end-steps are displayed too. (b) Relative energy-barriers and corresponding OOP polarization for sliding paths connecting hBN tetralayer's ABAC and ABCA.}
\label{fig3_path}
\end{figure}

HBN excels as an insulating and encapsulating material, while hGaN is a promising active material for optoelectronic and piezoelectric applications \cite{park_hexagonal_2023}; in both cases, the characterization of energy barriers and electric polarization within a slidetronics framework is of great interest. The inset in Figure \ref{fig2_sliding} shows the same ABCB-ABAB sliding path used to analyze hBN, but in this case for hexagonal GaN. This path shows a plateau-like polarization between such stackings, which mimics the hBN case. The energy barriers are larger (up to six times) for all the evaluated paths, and the magnitude of the polarization is $\sim3.5$ times larger in the hGaN case. This increase is mainly due to the presence of the Ga atom, with a larger number of electrons and less electronegative, which enhances the interaction between layers, allowing charge migration or distortion (different radii) of the p$_z$ orbitals, which favors asymmetry between stacked layers. A symptom of these more stressed GaN layers is the slightly less plane plateau in Figure \ref{fig2_sliding} (and the corresponding energy minima less defined), which is easily smoothed by tensile strains as NEB stabilizes hexagonal lattices slightly off local energy minima for a given stacking. This strengthening of the interaction is also reflected in the increase in energy barriers \cite{Wang}.

\subsection{Sliding Switched Polarization}
\label{Energie Barrier and Pathway}

In slidetronics, the ferroelectricity is switchable upon interlayer sliding driven by, e.g., a vertical electric field. Usually, non-equivalent stackings are connected through that process. However, sliding ferroelectrics enable high-speed data writing with low energy consumption while still ensuring robustness to thermal fluctuations \cite{fan__recent_2023}. The number of layers involved in the sliding of the tetralayer could vary, for example, in the transition from the ABAC configuration to ABCA two adjacent layers could be shifted simultaneously, as Figure \ref{fig3_path} shows. Studies on InSe bilayers suggest that this mechanism involves continuous sliding of the layers and achieves the reversal of the polarization sign~\cite{Sui2024}. In our case, the third and fourth layers move oppositely to reach the final configuration, resulting in a maximum energy barrier of about $\sim30\text{meV/f.u.}$, which is three times higher compared to the barriers observed on the left side of Figure \ref{fig2_sliding} (ABAC-ABAB-ABCB-ABCA). However, this barrier is approximately the same as the one for the ABCA-ABBA-ABBC-ABCA path. Figure \ref{fig3_path} allows us to compare the relative energy difference and the change in polarization of this double sliding with the paths previously evaluated in \ref{fig2_sliding}, which use one shifting layer. Another way to achieve similar changes in polarization, which could be more desirable from an application standpoint with respect to the comparison with pure electric-field-driven switching, by using two shifting layers, is by constructing pathways that allow the stacking to be tuned in a cyclic manner, i.e., from ABCA to an anti-cyclic order ACBA. In this case, the polarization value in hBN goes from -5.88 to 5.88 pC/m, resulting in the largest difference reported for sliding FE in hBN-based layered systems. 
\begin{figure}[t]
\centering
\includegraphics[width=1.0\columnwidth]{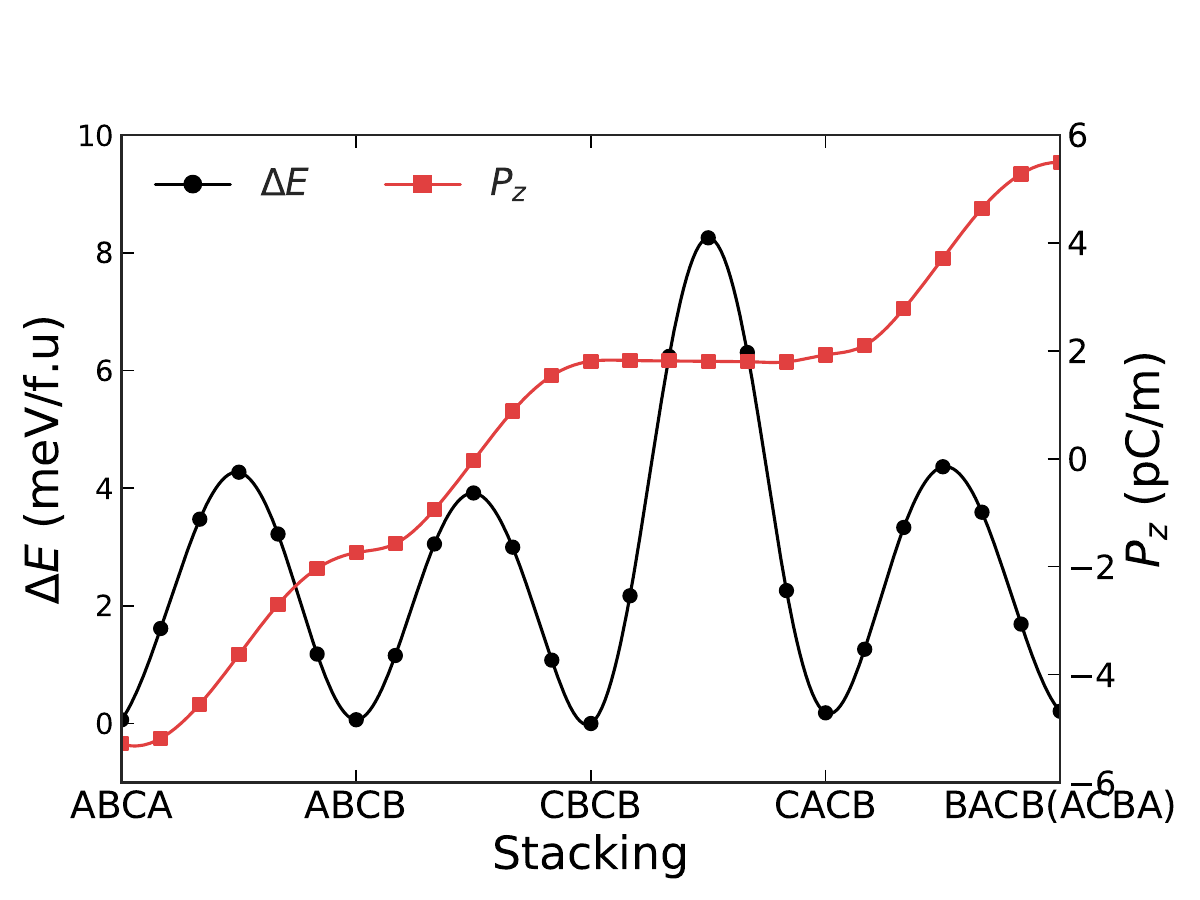}
\captionsetup{justification=raggedright, singlelinecheck=false, font=small}
\vspace{-0.0cm}
\caption{\justifying Relative energy-barriers and corresponding OOP polarization for a cyclic pathway from ABCA to ACBA. Largest $\left | P_{z} \right |=5.8 \text{pC/m}$, and largest $\left | \Delta E \right |\sim 10$.}
\label{fig5_path}
\end{figure}

\begin{figure}[t]
\centering
\includegraphics[width=1.0\columnwidth]{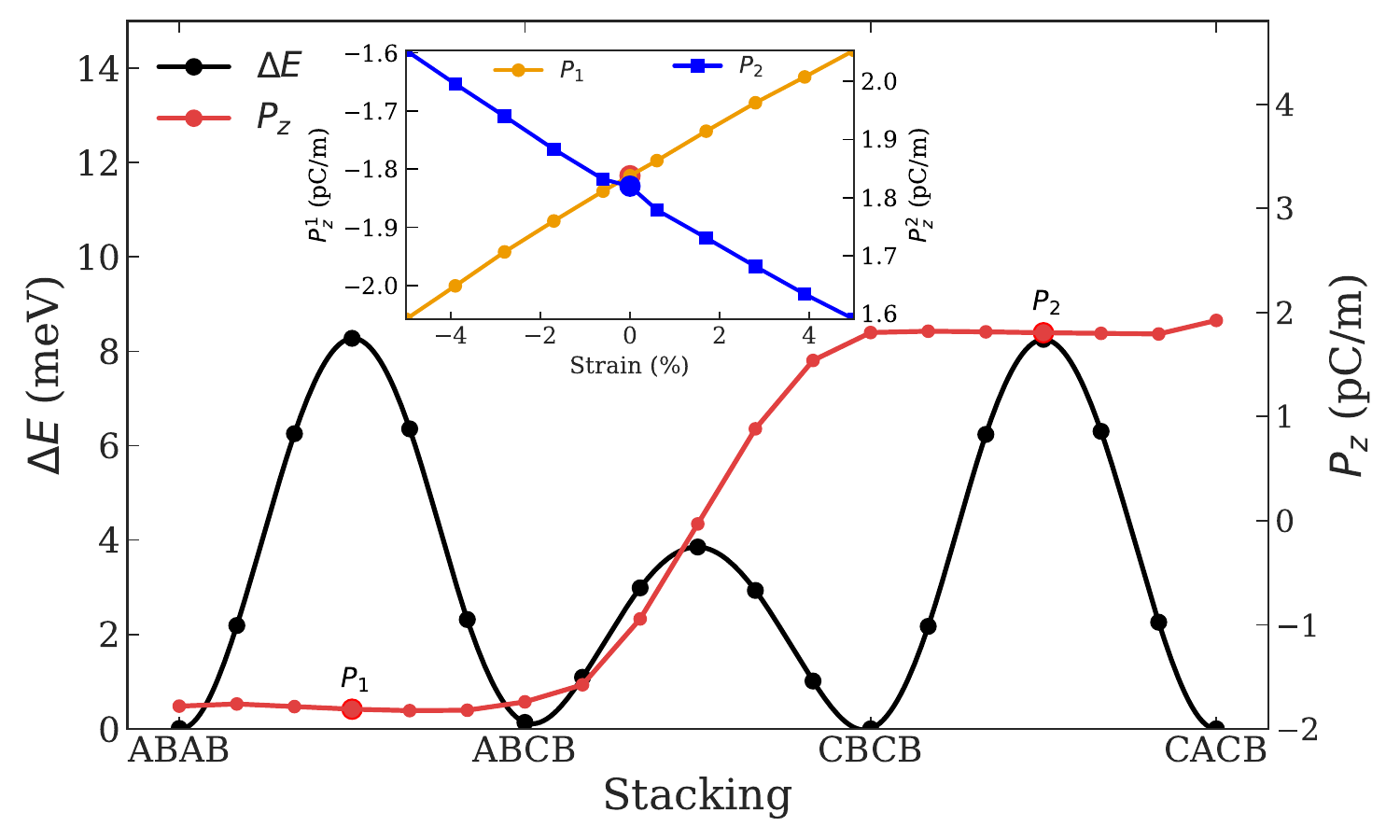}
\captionsetup{justification=raggedright, singlelinecheck=false, font=small}
\caption{\justifying Relative energy-barriers and corresponding OOP polarization for sliding paths connecting ABAB-ABCB and CBCB-ABCB. Inset: OOP polarization for in-plane strained tetralayers stacked ($0\%$) with structures corresponding to $P_{1,2}$ intermediate points signaled in the sliding path.}
\label{fig6_path_stable}
\end{figure}

Considering that the lowest energy barriers are obtained by moving one layer a distance \(d = \frac{1}{3}\) in the (1,1,0) direction, a pathway was designed to yield a landscape of such barriers. This path involves three sliding of one interface and one with two interfaces, i.e., ABCA-ABCB-CBCB-CACB-BACB. (BACB is equivalent to ACBA). Figure \ref{fig5_path} shows the energy barriers, with four processes of approximately 5 meV/f.u. and one of 10 meV/f.u., consistent with those mentioned previously. Specifically, sliding one interface yields barriers on the order of 4.5 meV, while two-interface sliding results in a double energy barrier value. Figures \ref{fig2_sliding} and Figure \ref{fig5_path} show electric polarization plateaus throughout the sliding processes that connect the ABAB and ABCB stacks as well as the CBCB and CACB stacks, respectively. Those sliding conserve in-plane translational symmetry as well as intralayer mirror symmetry, as can be seen in the inset of Figure \ref{fig6_path_stable} where in-plane strained layers respond with linear-like asymmetric FE, with respect to the plateau polarization values. This figure shows another path to flipping the polarization using these stackings. Due to the stability of the plateaus and the bipolar states at ABCB and CBCB, also within the plateau regions, the switching between positive and negative polarization could be performed by sliding between those ABCB and CBCB minima, which are separated by a small energy barrier. 

This is not a field-driven switching, therefore, when the plateaus are imminently reached along a slinding, as they do not obey to a process that needs a saturation field, as in magnetism, the FE can flips between polar states in a cheaper way, similar to negligible coercive field in magnetism. The three barriers in Figure \ref{fig6_path_stable} are among the smallest ones compared to the other sliding paths proposed above; therefore, ABAB and CACB are also plausible initial states for sliding-tuned switching since the polarization plateaus would guarantee that they have the same initial polarization as the ones in the internal minima; moreover, CBCB is equivalent to BABA, an arrangement that is the anti-cyclic counterpart of ABAB. Starting from the external minima mimics the preparation of a saturated state in magnetic hysteresis. The polarization plateaus are reminiscent of reports in hexagonal water ice, where temperature changes and applied electric fields can drive the compound into phases with constant polarization, attributed to a Kasteleyn transition \cite{Kasteleyn}.

\subsection{Bader Charge Analysis}
\label{Bader Charge Analysis}

Different experiments and calculations have attributed the origin of the polarization in layered hBN to either charge transfer processes or the distortion of the p$_z$ orbitals \cite{fan__recent_2023}. To gain further insight into the effects of such processes on the FE polarization, we used a Bader charges approach to estimate the polarization and then compare it with the DFT results.

\begin{figure}[t]
\centering
\includegraphics[width=1.0\columnwidth]{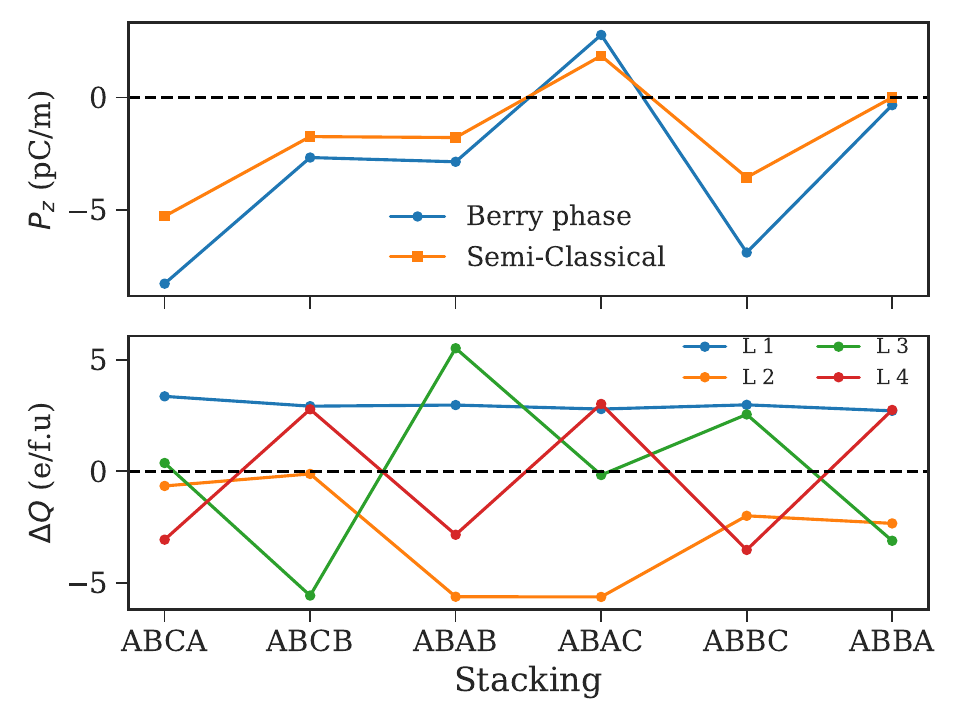}
\captionsetup{justification=raggedright, singlelinecheck=false, font=small}
\caption{\justifying (a) Comparison between the polarization calculated with the Berry phase method and the Bader-charges approach. (b) Excess/defect of charge in each layer with respect to an isolated hBN monolayer.}
\label{fig7_charge}
\end{figure}

Figure \ref{fig7_charge} displays the polarization calculated with these two different methods for different stacking. In terms of gain and loss of charge,
when a layer is changed between ABCA and ABCB, the fourth layer gains charge, while the third layer loses (layer C). The same pattern is observed in the paths ABAB to ABAC and ABBC to ABBA. With AB stacking in the two first layers, all translations of the third layer resulted in a charge gain in this layer and a compensated loss between the fourth and second layers. For example, on the path from ABCB to ABAB, the excess charge in the third layer varies from -5 to 5 me (milielectrons), while the sum of the excess charge in the second and fourth layers is -5 me. The second layer also displays charge losses that are staggered depending on the stacking of the third layer, i.e., for ABC we have the lowest value $\sim0$, for ABB an intermediate value $< 5$, and for ABA the largest value $> 5$.

Therefore, the first layer has a constant gain of charge across the stacking in Figure \ref{fig7_charge} (b), which might be useful considering that $L_{1}$ or $L_{4}$ should act as a buffer/substrate-like layer, and leaking processes must be avoided, for example, while scanning a biased tip above the tetralayer to promote sliding \cite{vizner_stern_interfacial_2021}. The second layer partially mimics these characteristics of $L_{2}$. Still, in a staggered-like way, suggesting that given an ABX trilayer arrangement, sliding an additional $L_{4}$ on top would always increase/flip the $\Delta Q$ if shifted in a cyclical manner A$\rightarrow$B$\rightarrow$C, and this seems to dominate the change of the polarization between the stacking defining each "step," that is, it would increase the polarization following the gains of charge of $L_{4}$. $L_{4}$ flips its $\Delta Q$ from negative to positive, as mentioned in the last case, but it also flips back $\Delta Q$ when this layer is fixed and $L_{3}$ is shifted in the same cyclical way A$\rightarrow$B$\rightarrow$C. The polarization in the last case decreases or stays approximately the same comparing the defining stacking, depending on how much charge $L_{3}$ has been able to balance with respect to the top layer change. The saw-like behavior for $L_{4}$ is very interesting, re-calling the difficulties that have always appeared in the differentiation of the stacking, for instance, in moir\'e domains \cite{CORTES2023111086}. The switch of $\Delta Q$ in the superficial layer, since $L_{1}$ is part of the substrate, could be used to determine a specific stacking given that a layer $L_{3}$ or $L_{4}$ is slid. In our tetralayer $L_{3}$ is the wildcard, its $\Delta Q$ opposes the $\Delta Q$ for $L_{4}$, while its magnitude balances the whole system similar to what happens in the trilayer case for the sandwiched layer \cite{CORTES2023111086}; the changes of charge in $L_{3}$ are larger when the $\Delta Q$ differences between two "steps" (sliding $L_{3}$) of the staggered-like $L_{2}$ charges are larger due to this balance.

In Figure \ref{fig7_charge} (a), for all stackings, there is an underestimation of the absolute value of the polarization by the semi-classical method, which is a symptom of the inability of the Bader method to adequately account for the slight tilt/deformation of the $p_{z}$ orbitals due to the different chemical pressure at each stacking. Polarization $P_{z}$ always increases/switches when $\Delta Q$ in $L_{4}$ increases/switches, which also happens at the "steps" of the staggered $\Delta Q$ for $L_{2}$; due to the balance mentioned above, polarization always increases when $\Delta Q$ decreases in $L_{3}$. Now, in terms of switching, ABAB-ABAC provides us with a symmetric one that depends only on manipulating the top layer, which seems more accessible experimentally. However, sliding such a layer on ABC provides us with a similar $\vert P_{z}\vert$ but without switching-like behavior, although in both cases, the second layer maintains a similar $\Delta Q$ between the end stacking. On the other hand, the larger change on $\vert P_{z}\vert$ is reached by sliding $L_{3}$, which also gives an asymmetric $P_{z}$ switching. If we look at Figure \ref{fig7_charge} and Figure \ref{fig6_path_stable}, we see that one of the main contributors to the plateaus is the balance of the transferred charge between the third, second, and fourth layers. Even small strains generate enough chemical pressure between the $p_{z}$ orbitals to increase/decrease the polarization, as illustrated in the inset Figure \ref{fig6_path_stable}. However, sliding a layer (third one) sandwiched between fixed-stacked layers, for example, between ABCB and ABAB, does not provide important polarization changes. Figure \ref{fig7_charge} suggests that this plateau behavior is possible mainly due to the charge balance trigger by the same layer that is sliding while the layers above and below gain/loss charge the same way, which is possible if they are stacked similarly between them, i.e., $L_{4}$ with $L_{2}$. If equivalent stacking is used after translation, cyclical permutation, mirror reflection, or inversion, as in the CBCB and CACB cases, where the third layer, in this case, would be read from right to left (sliding B$\leftrightarrow$A), we would expect a similar plateau behavior as suggested by Figure \ref{fig6_path_stable}. All this finally also suggests that a clever choice of fixed stacking $L_{1,2}$, according to the limitations/possibilities in which $L_{3}$ and $L_{4}$ can be experimentally manipulated, could provide us with a path to engineer such plateaus, which in turn could be used as switching states for FE purposes.

\subsection{Rotated Layers}
\label{Rotated Layers}
In the previous discussion, it has been clear that in the tetralayer sliding processes, manipulation of three layers ($L_{2}$ to $L_{4}$) could be enough to generate interesting FE responses. In that sense, let us now analyze and revisit the FE response of an hBN three-layer system, but in this case, adding a rotation. If we rotate an h-BN bilayer, a drastic reconstruction of the unit cell occurs for small rotation angles because of the high energy cost of the AA zones. Two equivalent triangular areas with stacks of AB and BA are observed, so net polarization is zero in the unit cell \cite{Jarillo-Herrero, Bennett2023}, due to the rotated bilayer space group (SG) P321 with three C$_2$ axes parallel to the plane of the layer, which prohibits a net vertical polarization.


\begin{figure}[t]
\centering
\includegraphics[width=1.0\columnwidth]{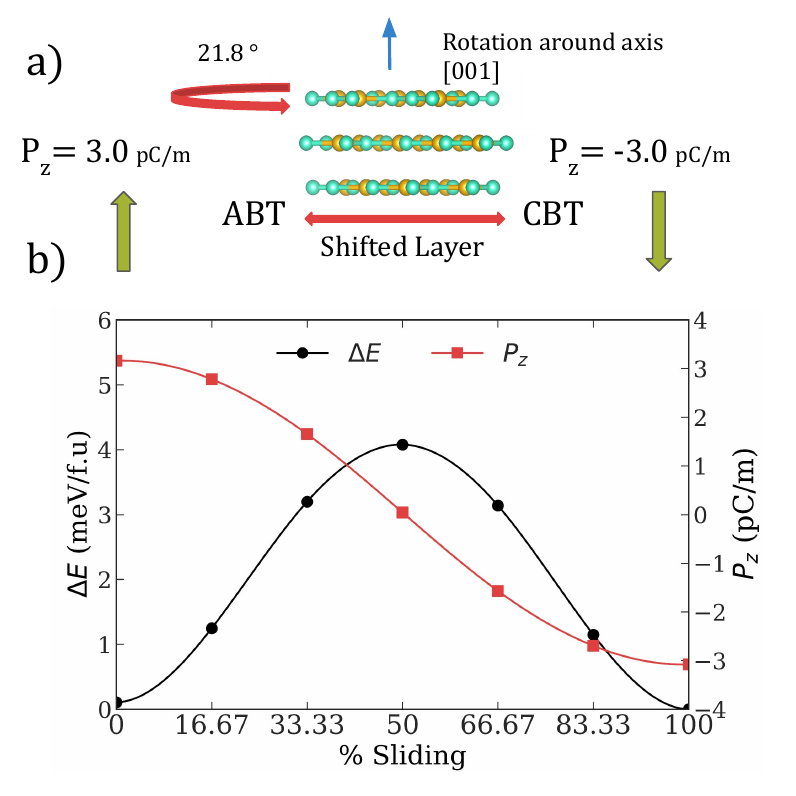}
\captionsetup{justification=raggedright, singlelinecheck=false, font=small}
\caption{\justifying (a) Illustration of the three-layer stacking with a rotated top layer. The sliding path ABT-CBT is performed using the bottom layer. (b) Relative energy-barriers and corresponding OOP polarization for the corresponding sliding path in panel (a).}
\label{fig8_path_ABT}
\end{figure}
However, for three or more layers, it is possible to obtain a net vertical polarization by rotating one or two of the layers. The unrotated ABA belongs to SG P-6m2, with inversion symmetry. Nonetheless, hBN trilayer with ABT stacking, that is, an AB stacking between the first and second layer and a rotated third layer, belongs to SG P3 with only  C$_3$ out of plane rotational axis \cite{Morell2011}. With a rotated middle layer, the ATA stacking will have no polarization due to the mirror ($m_z$) plane.

Let us consider ABT stacking, although a reverse stacking, such as TBA, would be equally valid and perhaps better from an experimental point of view.
First, we calculated the electric polarization for ABT and CBT stackings. Such polarization was obtained using the modern theory of polarization (Berry phase), as previously used in the sections above. Those two stacks, for a rotation angle of $21.8^{\circ}$, yielded results of $P_{z}=$ 3.0 and -3.01 pC/m, as shown in Figure \ref{fig8_path_ABT}. Second, we explore how to tailor the polarization between those two values. Sliding the twisted layer would have no effect;  we chose to slide the bottom (first) layer as it would be a superficial one, similar to the discussion of $L_{4}$ in previous sections. As Figure \ref{fig8_path_ABT} shows, sliding this bottom layer through a translation along the [110] axis, from ABT stacking to CBT, changes the polarization by smoothly switching its value from $P_{z}=$ 3.0 to -3.01 pC/m, respectively. The energy barrier along the pathway is approximately 4.07 meV / fu, which is comparable to the smallest barriers obtained for tetralayer, bilayer h-BN (5 meV), and smaller than previously reported barriers for multilayer MoS\textsubscript{2} (10-50 meV/f.u.), bilayer CrI\textsubscript{3} (23 meV/f.u.), and Cr\textsubscript{2}Ge\textsubscript{2}Te\textsubscript{6}, InSe \cite{ZHONG2023140430,Sui2024}. For this stacking, the rotated layer acts like an anchor, allowing the stabilization of the polar states between which the switching is performed; however, its influence on the height of the barrier seems comparatively negligible with respect to the previously discussed barriers. As discussed in the case of four layers, the number of interfaces appears to be one of the critical characteristics for layered hBN-based slidetronics.

\begin{figure}[t]
\centering
\includegraphics[width=1.0\columnwidth]{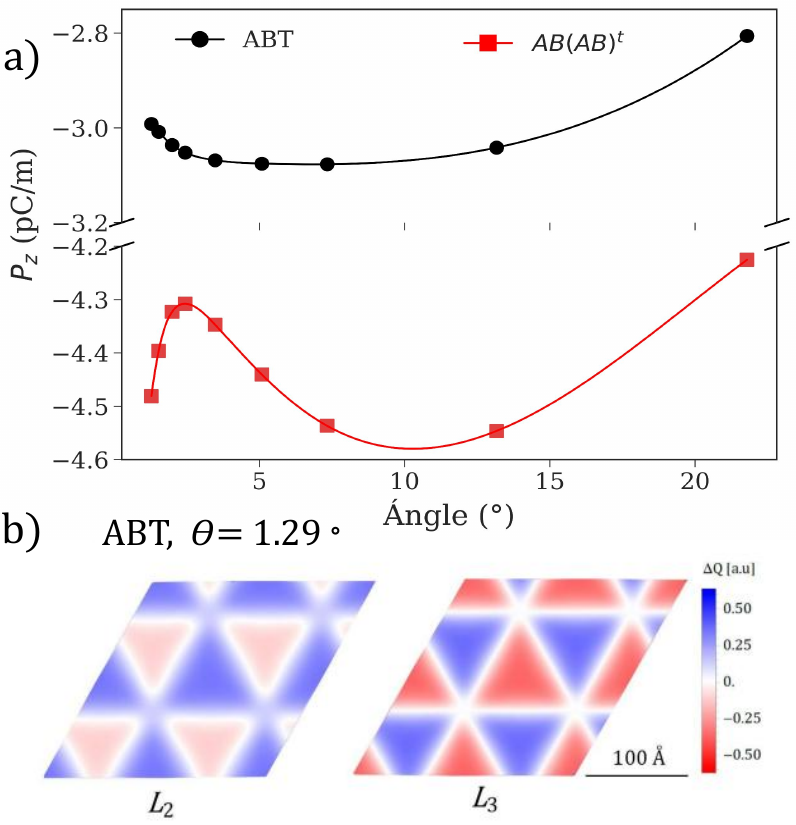}
\captionsetup{justification=raggedright, singlelinecheck=false, font=small}
\caption{\justifying (a) OOP as a function of the rotation angle $\theta$ in both trilayer (circles) and tetralayer (squares). (b) TB calculated $\Delta Q$ map in a 3x3 unit cell for $L_{2}$ and $L_{3}$ of the trilayer ABT, with $\theta=1.29^{o}$.
}
\label{fig9_polarization_angle}
\end{figure}

We explore now how the OOP polarization depends on the rotational angle. 
Using DFT, we can only calculate large rotation angles; therefore, we use TB modeling for small angles. This method has been successfully tested in the bilayer and trilayers \cite{CORTES2023111086}. In Table \ref{mi_tabla}, we show results for rotated angles of 21.8$^{o}$ and 13.2$^{o}$ in the case of the trilayer and tetralayer. The rotation is applied to the last two layers in the latter case. The Berry phase and the TB method yielded similar results. On the other hand, in Figure \ref{fig9_polarization_angle}(a), we show how the OOP values change with respect to the rotation angles in a trilayer ABT. Polarization values change less than 10\% in the entire range of angles.

We found the same triangular patterns associated with AB or BA stacking in trilayer rotated structures for low angles. This pattern is formed between the rotated layers, two and three; due to the reconstruction of the unit cell \cite{Yasuda,CORTES2023111086}. Figure \ref{fig9_polarization_angle} (b) shows the charge density for $L_{2}$ and $L_{3}$ in an ABT stack configuration, highlighting regions within the triangular patterns that indicate a net charge imbalance within the layers; the presence of the first layer leads to a net OOP. 

There are several ways to achieve polarization with rotated layers in four-layer structures. If we rotate only the outermost layer, for example, an ABAT, we would have the same scenario as with three layers. The situation changes when we rotate two layers. If we start from an ABBA stacking without net polarization and rotate layers three and four, we will not have a net polarization either (Twisted $AB(BA)^{t}$ belongs to SG P321). However, starting with a Bernal ABAB stacking and rotating the same two layers, we get a net OOP (SG P3) ($AB(AB)^{t}$).

Furthermore, in Table \ref{mi_tabla} DFT calculations were performed for two rotational angles of tetralayers $AB(AB)^{t}$, and $P_{z}$ was calculated with respect to the centrosymmetric unrotated stacking of ABBA. When $L_{2}$ and $L_{3}$ are rotated starting from ABAB stacking, a polarization of -4.58 pC/m is obtained, an increase of $\sim$ -2.0 pC/m compared to the unrotated. The value is similar for both rotated angles: 21.8$^{o}$ and 13.2$^{o}$. The TB Hamiltonian calculation for several rotational angles yields the same pattern; The polarization values, similar to trilayer ABTs, change slightly from 2$^{\circ}$ to 22$^{\circ}$ rotational angles.

From a symmetry analysis and after our calculations in Table \ref{mi_tabla}, it is corroborated that rotated bilayers generate no net polarization, only local polarizations: antiferroelectricity. Moreover, in three layers, if we rotate one of the outer layers, we will have a net polarization out of the plane; for four layers, if we rotate two of them, depending on the initial stacking, we could have a net polarization. With an initial stacking of AB(BA)$^T$, we will have an antiferroelectric system, and with AB(AB)$^T$, we will have a net OOP.

\begin{table}[t]
\caption{$P_{z}$ from DFT: stacking, rotated layer, SG.}
\centering
\begin{tabularx}{\columnwidth}{ Y Y Y Y Y }
\hline
Stacking & Rotated Layer & Space Group & Rotation Angle & $P_z$ (pC/m)  \\ \hline
ABAB & - & P3m1  & 0.00 & -2.50 \\
ABBA & - & P-6m2  & 0.00 & 0.00  \\
$AB(AB)^T$ & 3-4 & P3  & 21.8$^{\circ}$ & -4.58  \\
$AB(BA)^T$ & 3-4 & P321  & 21.8$^{\circ}$ & 0.00  \\
$AB(BA)^T$ & 3-4 & P321  & 13.2$^{\circ}$ & 0.00 \\
$AB(AB)^T$ & 3-4 & P3  & 13.2$^{\circ}$ & -4.58 \\
ABA & - & P-6m2  & 0.00 & 0.000 \\
$AB(A)^T$ & 3 & P3  & 21.8$^{\circ}$ & 3.00 \\
$CB(A)^T$ & 3 & P3  & 21.8$^{\circ}$ & -3.01 \\
\hline
\end{tabularx}
\label{mi_tabla}
\end{table}

\section{Conclusions}
\label{sec:conclusion}
In summary, this work shows ways to obtain spontaneous polarization in three and four hBN layers and pathways to change its sign.
Notably, tetralayer systems offer many spontaneous polarization states that can be achieved via sliding processes. Most of these processes have an energy barrier that suggests they could be reproducible at an experimental level. The energy barriers range from 5 to 30 meV/f.u. The lowest energy barriers are achieved when only one interface is involved in the sliding process. 

Different paths for changing polarization signs have been identified with energy barriers of $\sim$ 10 meV.
We also evaluated a simultaneous double sliding in  ABAC to ABCA through three pathways. The energy barrier is $\sim$ 30 meV. In the case of the ABAC-ABBC-ABBA-ABCA pathway, an energy barrier of 30 meV/f.u. was obtained. Generally, pathways with the lowest energy cost occur when single-layer sliding transitions are made from ABAC-ABAB-ABCB-ABCA, where the maximum energy barrier is $\sim$ 10 meV/f.u.

From the charge transfer analysis, it can be deduced that the charge transfer in L$_1$ is always the same, so it could be used as a sort of substrate for the other layers, while the change in $\Delta Q$ in the superficial layer could be used to determine a specific stacking given that a layer L$_3$ or L$_4$ is slid.

For some pathways, for example, between the ABAB-ABCB and CBCB-BACB stackings, the out-of-plane polarization remains constant, forming plateaus between different phases. The charge transfer analysis suggests that this plateau behavior is primarily possible due to the charge balance triggered by the same sliding layer, while the upper and lower layers gain/lose charge in the same way, which is possible if they are stacked similarly to each other, i.e., L\textsubscript{4} with L\textsubscript{2} (CBCB and CACB). All this finally also suggests that a clever choice of fixed stacking L\textsubscript{1,2}, according to the limitations/possibilities in which L\textsubscript{3} and L\textsubscript{4} can be experimentally manipulated, could provide us with a path to engineer such plateaus, which in turn could be used as switching states for FE purposes. However, this plateau is not robust under the application of strain, as even small pressure changes can significantly alter the out-of-plane polarization.

We also studied systems with rotated three- and four-layer structures, observing a net out-of-plane polarization in both cases. The polarization varies slightly within the range of evaluated rotational angles (from 2$^{\circ}$ to 22$^{\circ}$), with a variation of less than 10\%. In the trilayer with the third layer rotated ABT, it is possible to reverse the sign of the polarization by sliding the first layer, changing the polarization from 3.0 to -3.0 pC/m, with an energy barrier of approximately 4.07 meV/f.u.

Although our work focused primarily on hBN, calculations on similar nitrides such as hGaN suggest that the phenomenon is universal in this type of polar structure.
The energy barriers encountered suggest the feasibility of fabricating devices based on these structures.

\begin{acknowledgments}
The authors acknowledge financial support from ANID Chile, Fondecyt 1221301. JMF and ESM also thanks Proyecto Interno PI$\_$LII$\_$23$\_$09 USM.
\end{acknowledgments}


\bibliographystyle{vancouver}



\end{document}